\documentclass[journal=jacsat,manuscript=article]{achemso}

\usepackage{chemformula} 
\usepackage[T1]{fontenc} 
\usepackage{graphicx} 
\usepackage{xcolor}
\author{M. R. M. Atalla}
\affiliation{Department of Engineering Physics, \'Ecole Polytechnique de Montr\'eal, C.P. 6079, Succ. Centre-Ville, Montr\'eal, Qu\'ebec, Canada H3C 3A7}
\author{J. B\'elec}
\affiliation{Department of Engineering Physics, \'Ecole Polytechnique de Montr\'eal, C.P. 6079, Succ. Centre-Ville, Montr\'eal, Qu\'ebec, Canada H3C 3A7}
\author{E. Rahier}
\affiliation{Department of Engineering Physics, \'Ecole Polytechnique de Montr\'eal, C.P. 6079, Succ. Centre-Ville, Montr\'eal, Qu\'ebec, Canada H3C 3A7}
\author{S. Koelling}
\affiliation{Department of Engineering Physics, \'Ecole Polytechnique de Montr\'eal, C.P. 6079, Succ. Centre-Ville, Montr\'eal, Qu\'ebec, Canada H3C 3A7}
\author{K. Omambac}
\affiliation{Department of Engineering Physics, \'Ecole Polytechnique de Montr\'eal, C.P. 6079, Succ. Centre-Ville, Montr\'eal, Qu\'ebec, Canada H3C 3A7}
\author{P. Daoust}
\affiliation{Department of Engineering Physics, \'Ecole Polytechnique de Montr\'eal, C.P. 6079, Succ. Centre-Ville, Montr\'eal, Qu\'ebec, Canada H3C 3A7}
\author{S. Assali}
\affiliation{Department of Engineering Physics, \'Ecole Polytechnique de Montr\'eal, C.P. 6079, Succ. Centre-Ville, Montr\'eal, Qu\'ebec, Canada H3C 3A7}
\author{O. Moutanabbir}
\email{oussama.moutanabbir@polymtl.ca}
\affiliation{Department of Engineering Physics, \'Ecole Polytechnique de Montr\'eal, C.P. 6079, Succ. Centre-Ville, Montr\'eal, Qu\'ebec, Canada H3C 3A7}
\title {Extending Silicon Avalanche Photodetection Beyond 2 µm by Direct GeSn Integration}

\begin{document}


\begin{abstract}

Silicon avalanche photodiodes provide a technologically mature platform for sensitive photodetection, but their spectral response is intrinsically limited by the silicon bandgap. Extending their operation into the infrared requires the integration of narrow-bandgap absorbers while preserving efficient avalanche multiplication and compatibility with silicon processing. Here, we propose and demonstrate a monolithic approach that combines direct, buffer-free growth of GeSn on silicon with a lateral thin-junction separate-absorption-multiplication architecture. The GeSn layer, with a Sn composition reaching 6 at.\%, extends optical absorption to a wavelength of $2.6~\mu$m, while avalanche multiplication is spatially confined to an ion-implanted silicon lateral junction. This separation enables independent control of infrared absorption and carrier multiplication without the thick Ge virtual substrates conventionally used for GeSn epitaxy. GeSn-on-Si avalanche photodiodes exhibit low pre-breakdown dark current, stable breakdown at 72~V independent of device diameter, and clear infrared photoresponse extending beyond $2~\mu$m. At 78~K, the devices exhibit external quantum efficiency exceeding 100\%, reaching 163\% at $1.55~\mu$m, providing direct evidence of avalanche multiplication. These results establish direct integration of narrow-bandgap group-IV absorbers with silicon multiplication regions as a scalable strategy for extending the spectral reach of silicon avalanche photodetection, opening a route toward monolithic infrared detectors for sensing, imaging, communications, LiDAR, and quantum photonics.

\end{abstract}

\section{Introduction}

The infrared spectral region beyond $2\,\mu$m is a key technological frontier for next-generation infrared imaging, sensing, long-reach optical communication, and quantum systems\cite{meyer2025,kan2024near,shi2016transmission,jung2024low,zhang2026large,chow2026optical,eso2024impact}. In particular, the wavelength window around $2.1\,\mu$m exhibits lower atmospheric scattering, resilience to solar blindness and to poor weather, and offers enhanced eye safety due to reduced retinal exposure\cite{Dada2021near,eso2024impact,jiao2017ocular,li2021wavelength,panthier2022laser}. These attributes have stimulated a growing interest in applications such as LiDAR, biomedical diagnostics, environmental and industrial monitoring, quantum key distribution, night vision, and surveillance\cite{jiang2023long,li2023development,sun2026methane,altenburg2022situ,lough2020correlation,sun2026towards,nenonen2025high}. As demand for compact and deployable infrared systems grows, the development of high-performance photodetectors utilizing a scalable semiconductor platform has become a critical challenge.

To date, most detector technologies operating beyond $2\,\mu$m spectral range rely on III-V and II-VI compound semiconductors, including InGaAs, InGaAs/GaAsSb, AlInAsSb, and HgCdTe \cite{tosi2014low,jung2024low,liu2026high,anderson2022recent}. Although these material systems have enabled remarkable device performance, their adoption is often accompanied by challenges related to material cost, heterogeneous integration, manufacturing complexity, and limited compatibility with mainstream silicon (Si) microelectronics\cite{miao2023review, zhang2026overview}. Consequently, there remains a strong need for detector technologies that combine low noise with high sensitivity and compatibility with scalable manufacturing\cite{miao2023review,campbell2015recent}.

Although Si possesses many unique properties that make it an outstanding avalanche photodetector (APD), its bandgap limits its spectral operation to a wavelength of $1.1 \,\mu$m \cite{liu2026engineered,campbell2015recent,rochas2002low}. The mature Si ecosystem offers unmatched advantages in wafer-scale fabrication, process reproducibility, yield, and integration with complementary metal-oxide-semiconductor (CMOS) electronics\cite{modak2025cmos}. Thus, the development of APDs that are operational deeper in the infrared using Si-compatible semiconductors is highly coveted\cite{kang2009monolithic,xiang2022high,wang2022high,shi2024avalanche}. In this regard, GeSn alloys have emerged as a promising system owing to their tunable bandgap, compatibility with Si processing, and spectral coverage extending from the near-infrared to the mid-infrared\cite{reboud2024advances,moutanabbir2021mono,atalla2021all,tran2019si}. Indeed, several GeSn-based optoelectronic devices have been recently introduced in a variety of applications, including thermophotovoltaics, high-bandwidth photodetectors, light-emitting diodes, and lasers, which are essential building blocks for imaging, sensing, and optical communication systems\cite{atalla2022high,atalla2023extended,daligou2023group,luo2022extended,lemieux2025waveguide,atalla2024extended,lemieuxleduc2025transferprinted,luo2024mid,saleem2024infrared,cai2026polarization,kim2022enhanced,kim2023short}. 

Currently, many attempts have been made to achieve high-performance GeSn-based APDs\cite{miao2023review,finazzi2024modeling}. The vast majority of these devices reported to date rely, however, on growth on relaxed germanium (Ge) virtual substrates to accommodate lattice mismatch and facilitate high-quality GeSn epitaxy\cite{dong2014germanium,wanitzek2025gesn,rudie2025ge0,thai2026germanium}. The presence of these thick defective Ge buffer makes the control of the electrical field in GeSn-based APDs rather complex, yielding devices that suffer large leakage and premature breakdown\cite{finazzi2024modeling,giunto2024defects,julsgaard2020carrier,atalla2023dark}. As a matter of fact, GeSn-based APDs face significant performance constraints, specifically high dark current and modest multiplication factors\cite{assali2019enhanced,liu2022sn,buca2022room}. Herein, direct growth of GeSn on Si and lateral thin-junction APD design are proposed, demonstrated, and characterized. The obtained GeSn-on-Si APDs have a wavelength cut-off of $2.6 \,\mu$m  and show low dark current and EQE exceeding $100$\%, providing strong evidence of avalanche multiplication.


\medskip

\begin{figure*}[t]
    \centering
    \includegraphics[scale=0.45]{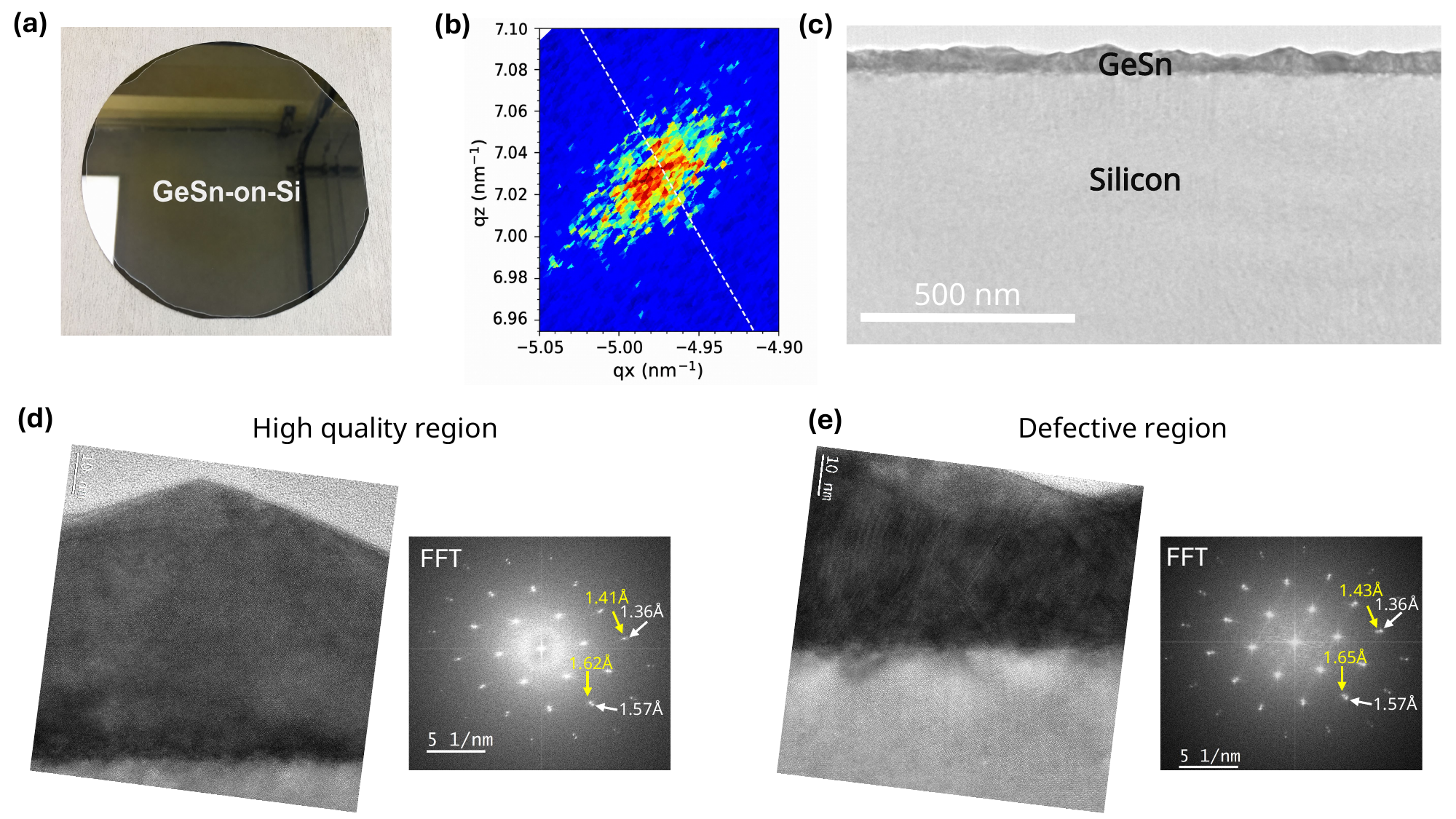}
    \caption{{\bf Material growth and characterization.} (a) Digital image of an as-grown 4-inch GeSn-on-Si wafer. (b) XRD-RSM around asymmetric $(\bar{2}\bar{2}4)$ peak of the as-grown GeSn-on-Si. (c) Cross-section TEM micrograph of an as-grown GeSn-on-Si layer. High-resolution TEM images of representative defect-free (d) and defective (e) regions of the as-grown GeSn-on-Si.}
\end{figure*}

\medskip

\section{Results and discussion}

\noindent {\bf Growth and characterization of GeSn-on-Si epilayers}

GeSn layers were grown directly on 4-inch Si (100) wafers in a low-pressure chemical vapor deposition (CVD) reactor using ultra-pure H\textsubscript{2 }carrier gas, and 10 $\%$ monogermane and tin-tetrachloride precursors. Before growth, the Si wafers were cleaned in a diluted HF solution, followed by washing with di-ionized water and  annealing in H\textsubscript{2} above 800°C. Typically, the large lattice mismatch between GeSn and Si is a limiting factor in controlling the growth, leading to frequent epitaxial breakdown and highly defective or amorphous GeSn. In this work, the optimal growth of GeSn layers was achieved at a temperature of 340 °C for a growth time of 20 min, a Ge/Sn ratio around 1500,  and a reactor pressure of 50 Torr. Fig. 1(a) displays a digital image of a GeSn-on-Si wafer right after epitaxial growth is complete. The mirror-like surface is a strong qualitative indicator of successful epitaxial growth, reflecting the absence of significant surface irregularities and the formation of a GeSn layer with highly uniform morphology and roughness well below the wavelength of visible light. Such a specular appearance is typically associated with stable two-dimensional growth and indicates the absence of significant surface defects and large three-dimensional islands.  

X-ray diffraction reciprocal space mapping (XRD-RSM), recorded around the asymmetric $(\bar{2}\bar{2}4)$ reflection, was performed to assess the structural quality and strain state of the grown layers. Fig. 1(b) shows a representative map. Notably, the Ge peak, typically observed around ($q_x = -4.99$ $\text{nm}^{-1}$; $q_y = 7.08$ $\text{nm}^{-1}$) in Ge-buffered GeSn heterostructures, is absent, confirming that the GeSn layer was grown directly on the Si wafer without an intermediate Ge layer. In addition, the GeSn diffraction peak lies close to the relaxation line, indicating that the layer is largely strain-relaxed. The GeSn peak is relatively broad, consistent with the compositional gradient of 3 - 6 at.\% Sn. Cross-sectional transmission electron microscopy (TEM) confirmed the growth of a continuous GeSn film, but with an irregular microstructure and a thickness in the 50- 80 nm range (Fig. 1(c)). High-resolution TEM images revealed highly crystalline faceted regions (Fig. 1(d)), as well as regions containing a high density of dislocations (Fig. 1(e)). However, as shown below, these structural and compositional variations occur on a scale that does not affect the performance or device-to-device consistency of the APDs fabricated from the grown GeSn-on-Si material.

\medskip

\begin{figure*}[t]
    \centering
    \includegraphics[scale=0.54]{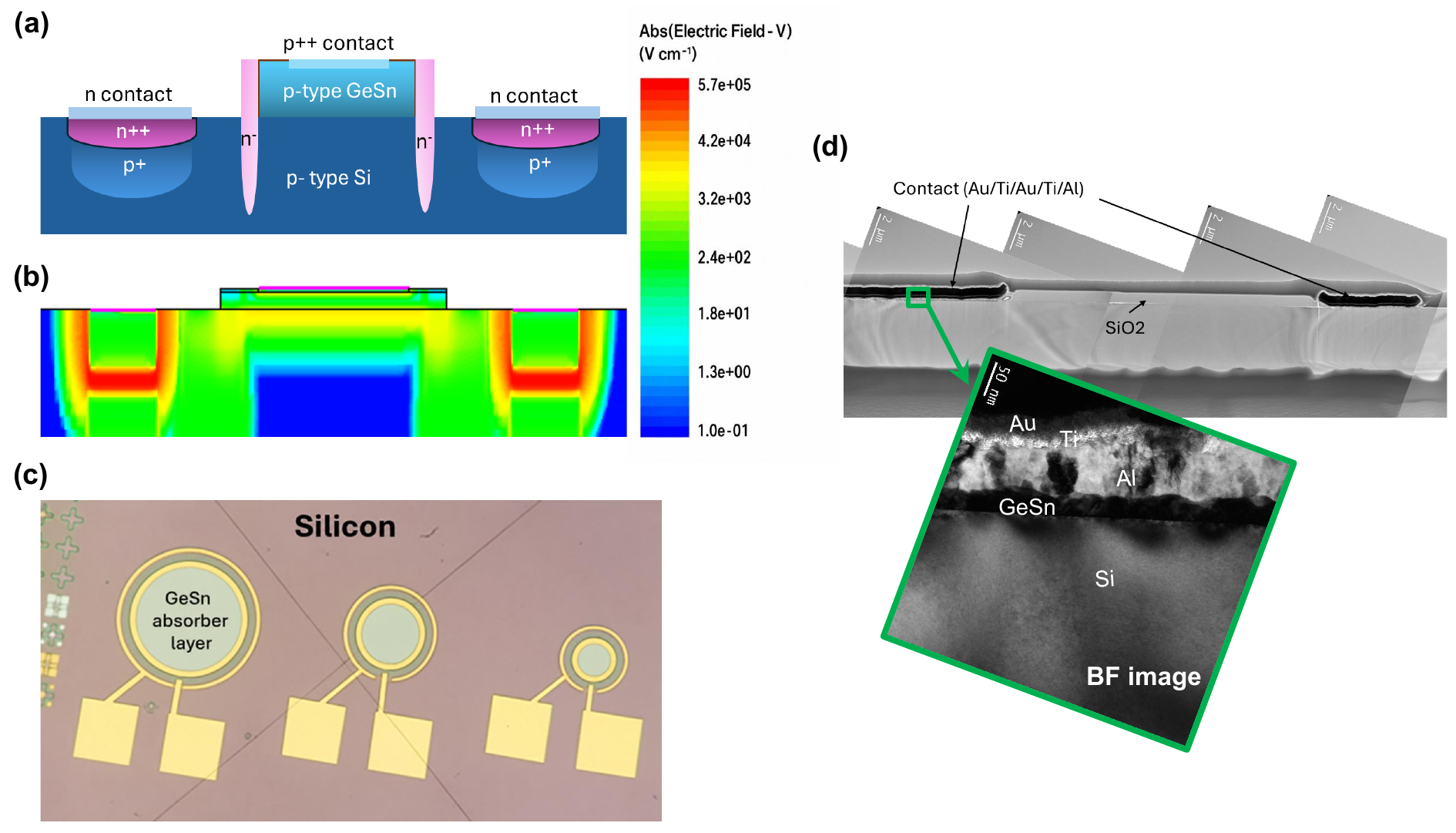}
    \caption{{\bf GeSn-on-Si APD design and fabrication.} (a) Schematic of the GeSn-on-Si APD design illustrating the guard ring and carrier multiplication regions. (b) The corresponding 2D electric field map calculated using Sentaurus TCAD, confirming the very high field in the Si substrate where avalanche multiplication takes place, and the electric field still propagates into the GeSn layer, which is useful for photo-carrier collection. (c) Micrograph of the fabricated and completed GeSn-on-Si APD device with various diameters in the $25-200 \,\mu$m range. The GeSn absorbing layers can be seen as the gray regions. (d) TEM cross-sectional images of a processed GeSn-on-Si device showing the GeSn layer and the metal contacts.}
\end{figure*}

\medskip

\noindent {\bf Lateral thin-junction GeSn-on-Si APD photodetectors}

An avalanche device that has a separate optically active absorbing region and carrier multiplication region requires a careful design to ensure optimal electric field distribution. Indeed, the field should be sufficiently high in the multiplication region, low in defect regions, and significant and not diminishing in the absorbing region. A lateral thin-junction with separate-absorption-multiplication (SAM) APD design is proposed herein. The thin-junction multiplication region provides control over the device breakdown voltage by optimizing the dopants concentrations and distribution in the pn-junction. It also provides a simpler access for the ion implantation beam to the designated multiplication region as compared to a vertical absorption multiplication region APD device. A schematic of the device architecture is shown in Fig. 2(a). The design includes a circular mesa of GeSn surrounded by lightly doped n-type deep wells to minimize the surface current that could potentially limit the device performance. The multiplication region is designed to be located in Si, laterally separated from the active GeSn mesa where the n-contact is atop a heavily doped n-type well, which is vertically situated over a p-type doped well. The various doping levels and types were achieved using ion implantation, where the p-type and n-type behaviors are obtained by implanting boron (B) and phosphorus (P) ions, respectively. As shown in Fig. 2(b), Sentaurus TCAD simulations were implemented to optimize the proposed design of GeSn-on-Si APD. One of the advantages of this device structure is that the absorbing layer can be relatively thin to allow for efficient propagation of the electric field from Si to the GeSn layer. Additionally, the multiplication region in Si underneath the n-type contact has a very high electric field. The field is not low within the GeSn region, yielding an efficient collection of photo-generated carriers. 
It is also worth mentioning that this lateral thin-junction APD design avoids the need for a high breakdown bias compared to a thick-junction APD that uses Si substrate as an i-layer and p-contact on the back side of the Si substrate\cite{liu2026engineered}.

The device fabrication started with patterning Ti/Au alignment marks using standard photolithography. Then, ICP RIE etching was utilized to form circular GeSn mesas with various diameters in the $25\text{--}200\,\mu$m range. These constitute the optically active regions of the fabricated devices. Ion implantation is performed using a tandem particle accelerator to dope different regions in Si, as indicated in Fig. 2(a), and the sequence of the ion implantation steps was as follows: Implantation of P ions at a dose of $2 \,\times\, 10^{12}$ cm$^{-2}$ at $170$ keV to obtain a lightly doped n-type deep well, implantation of P ions at a dose of $7 \,\times\, 10^{15}$ cm$^{-2}$ at $40$ keV to obtain a heavily doped shallow well for the n-type contact, implantation of the B ions at a dose of $2 \,\times\, 10^{13}$ cm$^{-2}$ at $120$ keV to obtain a moderately doped p-type well underneath the previously mentioned n-type well, and finally, implantation of B ions at a dose of $1.5 \,\times\, 10^{15}$ cm$^{-2}$ at $40$ keV to obtain a p-type contact using a heavily doped shallow well. A thermal annealing step was necessary to activate dopants and remove ion-induced damage, especially in the multiplication region of the device. It should be mentioned that, while Si samples could be annealed at temperatures as high as $900\,^{\circ}\text{C}$, the annealing temperature of the GeSn-on-Si devices is kept below $450\,^{\circ}\text{C}$ to avoid Sn segregation and preserve the integrity of the GeSn layer. 

To isolate electrode pads from doped regions and the surrounding undoped Si region, a  $600$ nm-thick passivation layer of SiO$_2$ was deposited using plasma-enhanced CVD. The SiO$_2$ film is patterned and wet-etched to form openings for the electrode metals for the n- and p-type contacts. Metal contacts consisting of Al/Ti/Au/Ti/Au were deposited at thicknesses of $80$ /$15$ /$85$ /$35$ /$170$ nm. Then, contact annealing is performed at $385\,^{\circ}\text{C}$ for further improvements, as shown in the SI (Fig. S2(c)).  

\medskip

\begin{figure*}[t]
    \centering
    \includegraphics[scale=0.7]{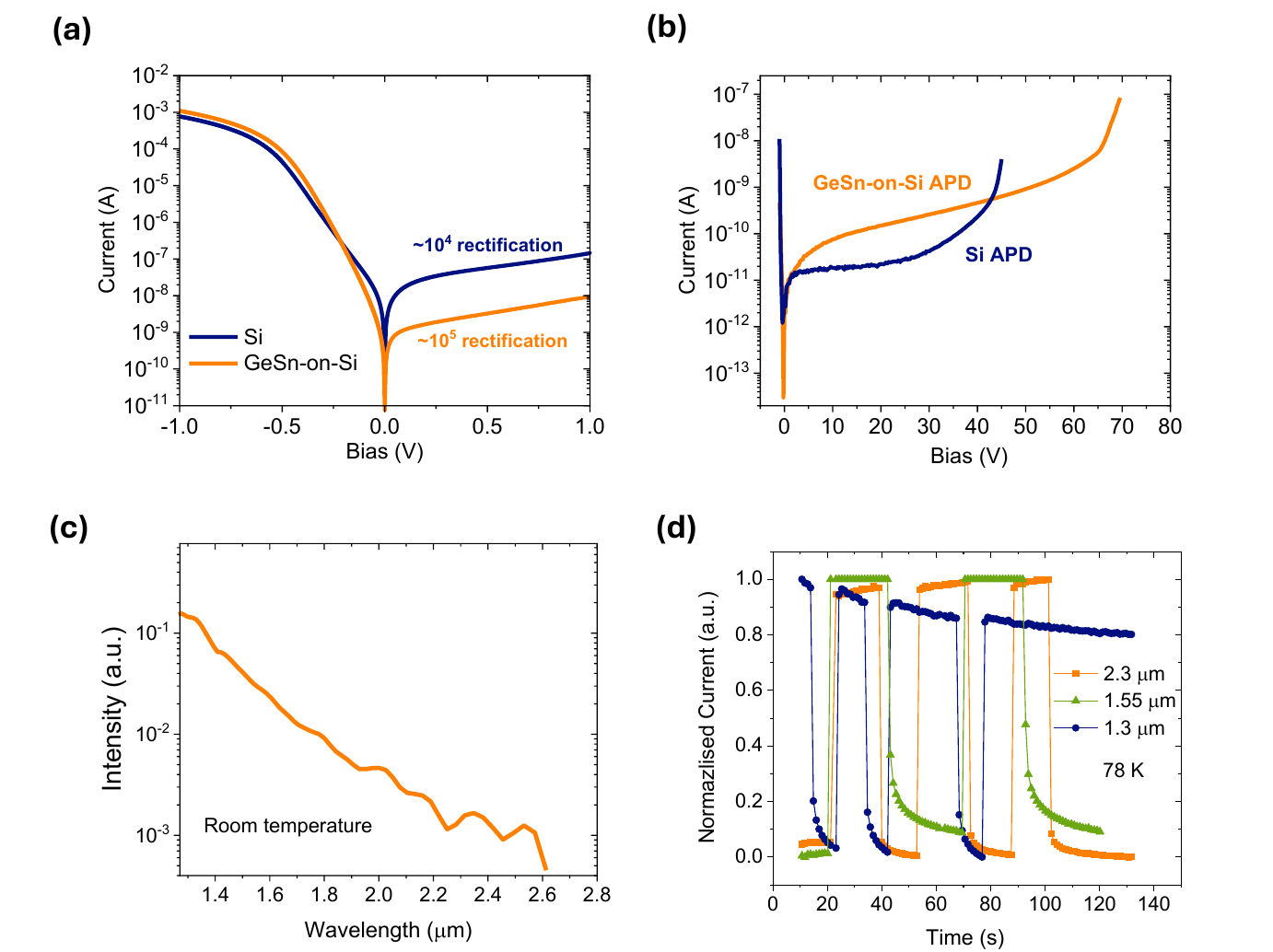}
    \caption{{\bf I-V characteristics of Si and GeSn-on-Si APDs and their spectral response.} (a) Room temperature I-V curves for the Si and GeSn-on-Si devices indicating a high rectification ratio at $1$ V. (b) Same as (a) but at $78$ K depicting the breakdown voltages of these two devices. (c) Spectral intensity of the GeSn-on-Si material as measured using an FTIR spectrometer at RT. (d) The normalized current as a function of time at $78$ K when the device is illuminated with a laser at wavelengths of $1.3$, $1.55$, and $2.3 \,\mu$m .}
\end{figure*}

\medskip

To discuss the device performance, we first examine the I-V curves in the dark. These analyses show a rectification ratio around $10^4$ and $10^5$ for the Si and GeSn-on-Si APDs, respectively, obtained at $0.5$ V for the $100\,\mu$m-diameter device (Fig. 3(a)), which indicates the relatively successful activation of dopants and annihilation of the lattice defects created during ion implantation. It is worth mentioning that the low dark current should significantly increase the detectivity of these devices compared to photoconductive devices that suffer from high dark current and high noise.\cite{atalla2021all,tran2019si} This would allow the operation of GeSn-on-Si APD devices without the need for a lock-in technique to extract the photocurrent signal.

Note that the different annealing temperatures are expected to affect the diffusion of dopants and damage recovery in the two sets of devices. This can be better observed when the reverse bias is increased up to device breakdown, as shown in Fig. 3(b). This figure displays the I-V curves of the Si and GeSn-on-Si APDs at $78$ K measured using a cryo-probe station in the dark. The Si reference APD maintains a low leakage current, whereas the GeSn-on-Si APD shows a larger increase in leakage current and breakdown voltage. Interestingly, both devices demonstrate a low reverse current below a few nA up to a bias of $1$ V, cooling down was necessary to prevent the devices from breaking down at much higher bias due to the increased leakage current. Similar I-V curves were obtained for other devices at different diameters, as shown in the SI (Figs. S2(a) and (b)).

To further probe the properties of the GeSn-on-Si APD, the spectral intensity of the GeSn-on-Si material was measured using an FTIR spectrometer, as shown in Fig. 3(c). The measurement was made at 1 V at room temperature, showing a monotonic decrease as the wavelength increases until it reaches a cutoff wavelength of $2.6\,\mu$m. This gradual decrease in responsivity at higher wavelengths is indicative of a compositional gradient in the GeSn layer, which is consistent with the measured Sn content reaching $6$ at.$\%$ (Fig. 1(b)). The evidence of the effective contribution of the GeSn layer to the total current of the GeSn-on-Si device under illumination is the significant photo-current generation at $78$ K when lasers at wavelengths of $1.3$, $1.55$, and $2.3 \,\mu$m are incident on the device, as shown in Fig. 3(d).

\begin{figure*}[t]
    \centering
    \includegraphics[scale=0.65]{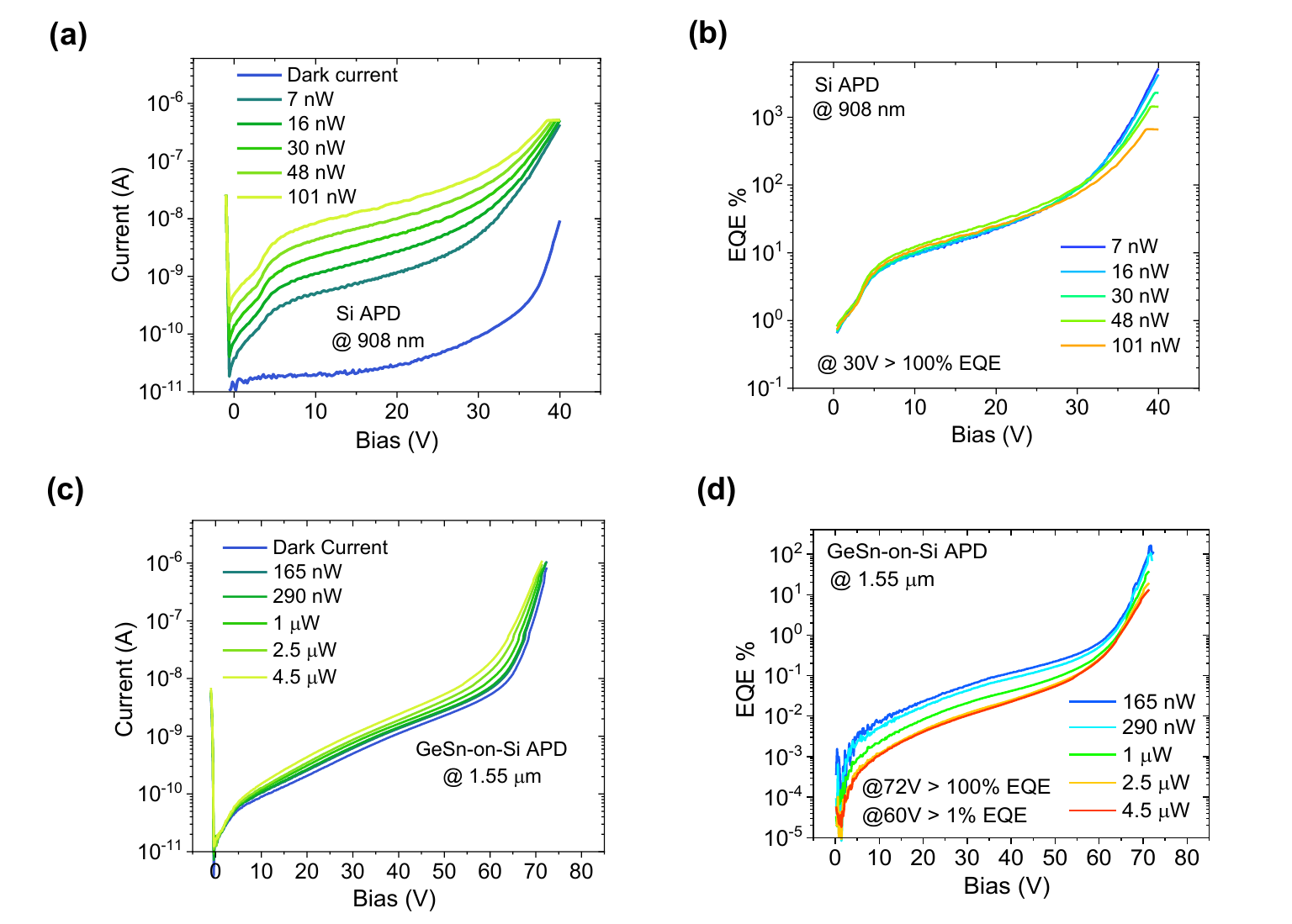}
    \caption{{\bf I-V characteristics and EQE of Si and GeSn APDs at various optical powers.} (a) I-V curves for the Si APD device measured at $78$ K under dark and under $908$ nm laser illumination at various powers. (b) EQE of the same device in (a) at the same temperature and optical powers. (c) I-V curves for the GeSn-on-Si APD device measured at $78$ K under dark and under $1.55\,\mu$m laser illumination at various powers. (d) EQE of the same device in (c) at the same temperature and optical powers.}
\end{figure*}

\medskip

\noindent {\bf EQE and avalanche characteristics of Si and GeSn-on-Si APD devices}

Note that after all ion implantation steps, the Si APD devices were annealed at a significantly higher temperature and for a longer duration than the GeSn-on-Si devices, whose metastable nature limits the processing temperature. Moreover,  the narrow bandgap absorbing layer in GeSn-on-Si APDs is associated with higher density of thermally generated carriers. Thus, a direct comparison between the two sets of devices must consider these differences. Fig. 4(a) displays the I-V curves of the Si APD device of diameter $100\,\mu$m measured at $78$ K in the dark and under illumination conditions. The current remains below $40$ pA until the reverse bias reaches $20$ V, indicating the very low leakage in these Si devices. Afterward, the current increases exponentially until the device reaches breakdown at $41$ V. It should be noted that the breakdown voltages of these Si APDs were found to be independent of the device diameter, as shown in the SI (Fig. S2(a)), indicating that they are not influenced by the geometry of the device but by the high electric field induced in the multiplication region inside the Si material. The total current measured under a $908$ nm continuous-wave laser is also shown in Fig. 4(a). The total current is plotted at incident optical powers of $7$, $16$, $30$, $48$, and $101$ nW. Due to the remarkably low dark current, the total current is significantly higher than the dark current, and it increases drastically above $20$ V until breakdown in a behavior similar to that of the dark current. The external quantum efficiency (EQE) for this APD can be calculated as \cite{streetman2000solid}  
 
 \begin{equation}\label{EQEeq}
    EQE_{\lambda} = R_{\lambda} \times E_{ph} / e , 
\end{equation}

where $R(\lambda)$ is the responsivity at a wavelength $\lambda$, $E_{ph}$ is the photon energy at the same wavelength, and $e$ is the electron charge. The responsivity is calculated as $R(\lambda) = (I_{Total} - I_{dark})/P_{\lambda}$, where $I_{Total}$, $I_{dark}$, and $P_{\lambda}$ are the total current, the dark current, and the incident optical power at wavelength $\lambda$, respectively. In Fig. 4(b), EQE is plotted for the same device at the five different optical powers considered in Fig.4(a), in reverse bias. It can be readily seen that the EQE exceeds $100$\% as the reverse bias exceeds $30$ V, and it exceeds $1000$\% at $37$ V. It is important to point out that the EQE exceeding $100$\% indicates that the number of generated photo-carriers is more than the total number of incident photons on the device, which is strong evidence of carrier avalanche mulitpilication occurring in the high electric-field region in the Si APD \cite{You:23,tsang2020quantum,kamal2025advances}. Generally, the EQE curves in Fig. 4(b) coincide with each other regardless of the incident optical power, except at the highest optical power $101$ nW when the bias exceeds $32$ V. This is most likely because of an increase in series resistance, and it results in a slight sign of saturation in the photo-generation dominated current. 

\medskip

The fabricated GeSn-on-Si APD devices were then measured at $78$ K. In Fig. 4(c), it is observed that the dark current of a $100\,\mu$m diameter GeSn-on-Si device has rapidly increased to $1.12$ nA at $40$ V, and it continued to increase until it approached breakdown at a bias of $72$ V. The I-V curves of the GeSn-on-Si device are also presented under illumination of a continuous-wave $1.55\,\mu$m laser at optical powers of $0.165$, $0.29$, $1$, $2.5$, and $4.5\,\mu$W, showing a monotonic increase in the total current as the optical power increases. It is important to mention that the GeSn-on-Si devices have demonstrated a higher leakage current and a higher breakdown voltage compared to those of the Si devices, most likely because of the low post-ion implantation annealing temperature compared to that of the Si devices. Despite the relatively high dark current, the total current is higher than the dark current at relatively low optical power, and it increases nonlinearly above $60$ V until breakdown in a similar behavior to that of the dark current. Similarly to Si reference devices, the breakdown voltages of these GeSn-on-Si APDs are also found to be independent of the device diameter, which indicates that they are influenced by the high electric field induced in the multiplication region inside the Si substrate, as shown in the SI (Fig. S2(b)). 

The EQE for this GeSn-on-Si APD has been evaluated and plotted in Fig. 4(d) as a function of reverse bias for the same device at the same five different optical powers considered in Fig.4(c). We note that EQE exceeds $1$\% at a bias above $60$ V where a breakdown knee starts, and it eventually reaches $163$\% at a breakdown bias of $72$ V for the optical power of $165$ nW. It is important to point out again that EQE exceeding $100$\% indicates that the number of generated photo-carriers is more than the total number of incident photons on the device, confirming that carrier avalanche multiplication takies place in the high electric-field region of the GeSn-on-Si APD. 

It is evident from Fig. 4(d) that the EQE curves do not coincide with each other, in sharp contrast with the Si APD case (Fig. 4(b)). While the Si device utilizes the entire Si bulk substrate to absorb the incident light, the GeSn-on-Si APD has an active material thickness of only $50-80$ nm to absorb the $1.55\,\mu$m laser. This is likely the reason behind the reduced EQE as the incident optical power increases, and it strongly indicates a sign of saturation in the total current that is dominated by photo-generated carriers. Similar I-V and EQE characteristics were obtained at $78$ K using a $1.3\,\mu$m continuous wave laser, as shown in the SI (Fig. S1), and the EQE exceeds $100$\%, confirming the avalanche properties of these GeSn-on-Si lateral APDs. Additionally, as mentioned above, the on/off normalized total current as a function of time is exhibited in Fig. 3(c) and reveals a slightly slow decay when light is off, which indicates the relatively low bandwidth of these devices and imposes limitations on the single-photon counting capability of the APD devices, as detailed in the SI (Fig. S3).

\section{Conclusion}

This work demonstrates monolithically integrated GeSn-on-Si APDs with a cutoff wavelength of $2.6\mu$m, enabled by the direct CVD growth of GeSn on Si without an intermediate Ge virtual substrate. This simplified epitaxial architecture facilitates the implementation a laterally separated absorption-multiplication design and Corbino geometry, allowing infrared absorption in GeSn while confining avalanche multiplication to Si. Despite the relatively thin ($\sim 50$--$80$ nm) and compositionally graded GeSn absorber, the devices exhibited excellent rectification, low pre-breakdown dark current, clear avalanche behavior, clear response under $2.33\mu$m illumination, and EQE exceeding $100\%$, reaching a maximum of $163\%$. The substantially higher EQE measured in the Si reference APDs indicates that the present performance is primarily limited by optical absorption in the thin GeSn layer rather than by the avalanche multiplication mechanism. These results establish direct GeSn-on-Si growth as a viable route toward all-group-IV APDs operating beyond the conventional wavelength range of Si and Ge technologies. Increasing the GeSn absorber thickness and Sn content, together with further optimization of the device architecture, should simultaneously enhance absorption, responsivity, and avalanche gain while extending the spectral response further into the infrared. This work therefore provides a pathway toward fully Si-compatible tunable infrared APDs for integrated imaging, spectroscopy, LiDAR, environmental sensing, and secure optical communication systems.

\noindent {\bf Methods}

\noindent {\bf GeSn on Si epitaxial growth abd characterization.} The GeSn layer was directly grown on a 4-inch Si (100) wafer in a chemical vapor deposition (CVD) reactor using monogermane (GeH$_4$) and tin-tetrachloride (SnCl$_4$) precursors while having an ultra-pure H$_2$ carrier gas. The growth of the GeSn layer was performed at a temperature below 350 °C for a variable growth time. The Sn composition of the layer was controlled by the growth temperature, while the Ge/Sn ratio in gas phase was kept constant to prevent phase separation during growth. TEM specimens were prepared using a focused ion beam (FIB) in a FEI Helios NanoLab~600 operating with a 30~keV Ga ion beam. TEM imaging was performed on a C-FEG JEOL JEM-F200 operating at 200~kV. The crystallinity of the as-grwon layers was investigated using high-resolution X-ray diffraction (XRD) in a Bruker D8~Discover system equipped with a Cu~K$\alpha_1$ source, a triple-bounce Ge(220) analyzer, and a Ge(220) monochromator. The ($\overline{2}\,\overline{2}\,4$) reflection was used for reciprocal space mapping (RSM) analysis.

\noindent {\bf Dark and photocurrent measurements.} The I-V measurements were acquired using a Keithley 4200A parameter analyzer connected to a probe station. The photocurrent was measured at $1.55\,\mu$m wavelength, showing a strong rectifying behavior regardless of the device diameter. Moreover, the spectral responsivity was measured using a Bruker Vertex 80 FTIR spectrometer. The IR light source of the spectrometer was incident on the GeSn device, and the electrical signal was measured using a Zurich Instruments lock-in amplifier that was locked to the frequency of a chopper in the light path of the light source. The lock-in signal is fed to the spectrometer electronics to eventually obtain the photocurrent as a function of wavelength. The low-temperature I-V measurements were done using a Janis cryo-probe station, and the light was injected onto the device via an optical fiber and a focusing parabolic mirror.

\noindent {\bf Device EQE measurements,} The devices were measured at low temperature, $78$ K,  in a Janis cryo-probe station, and continuous wave lasers were incident and focused on the active area of the device using an optical fiber focuser. The Thorlabs PM100D2 power meter was used to measure incident laser power down to the nW range at wavelengths $1.3$, $1.55$, $2.3\,\mu$m. A Keithley 4200A parameter analyzer was used to accurately measure the low-noise I-V curves of these group IV avalanche devices. Optimization of focused laser spot location was made to find the highest photocurrent at the beginning of the measurement and was kept fixed afterwards throughout the rest of the EQE measurements. while the laser spot diameter was $50\,\mu$m the device diameter was $100\,\mu$m which confines the light within the device.

\bigskip

\begin{acknowledgement}

The authors thank J. Bouchard for the technical support with the CVD system. O.M. acknowledges support from NSERC Canada (Discovery, SPG, and CRD Grants), Canada Research Chairs, Canada Foundation for Innovation, Mitacs, PRIMA Québec, and Defence Canada (Innovation for Defence Excellence and Security, IDEaS).

\end{acknowledgement}

\begin{suppinfo}
See supplement file for supporting content. 
The data that support the findings of this study are available from the corresponding author upon reasonable request.\\
\noindent The authors declare no conflicts of interest.

\end{suppinfo}

\bibliography{main}


\end{document}


\newpage

\subsection {\bf S1. EQE and avalanche characteristics of Si and GeSn-on-Si APD devices under $1.3 \,\mu$m laser illumination.}

The GeSn-on-Si APD sample was also measured using a Cryo-probe station at $78$ K in the dark and under illumination at various optical laser powers at a wavelength $1.3\,\mu$m, as shown in Fig. S1(a). Similarly to Fig. 4(c) the I-V curves under illumination depict a high leakage current and a breakdown bias around $72$ V. This Fig. S1(a) under $1.3\,\mu$m illumination is additional evidence on the functional contribution of the GeSn layer to the total current of the device owing to a strong photo-carrier generation under the influence of the incident NIR photons, which rules-out the contribution of Si to the photocurrent as the incident photons have energy smaller than Si bandgap. In Fig. 4(c), the I-V curves of the GeSn-on-Si device under illumination of a continuous wave $1.3\,\mu$m laser at optical powers of $0.138$, $0.293$, $0.745$, $3.4$, $9.1$, $17.1$ and $45\,\mu$W are presented, showing a monotonic increase in the total current as the optical power increases. Despite the relatively high dark current, the total current is higher than the dark current at relatively low optical power, and it increases nonlinearly above $60$ V until breakdown in a similar behavior to that of the dark current. 

\medskip

\begin{figure*}
    \centering
    \includegraphics[scale=0.75]{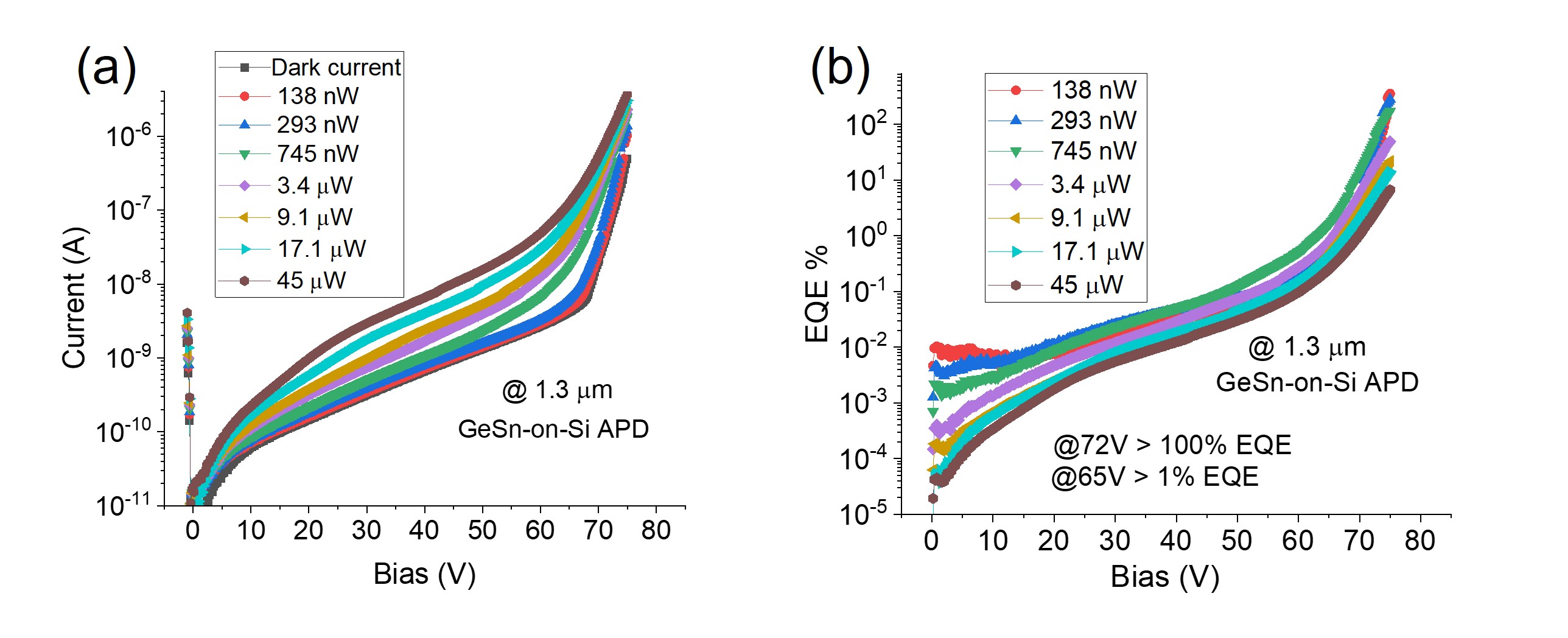}
    \caption{{\bf LT I-V and EQE of GeSn-on-Si APDs at various optical power $1.3\,\mu$m laser illumination.} (a) I-V curves for the GeSn-on-Si APD device measured at $78$ K in dark and under illumination. (b) EQE of the Same device in (a) at the same temperature and optical powers.}
\end{figure*}

\medskip

Similarly to Fig. 4(d), the EQE for this GeSn-on-Si APD has been calculated, and it is plotted, in Fig. S1(b), as a function of reverse bias for the same device and the same various optical powers considered in Fig. S1(a). It can be readily seen that the EQE eventually reaches $355$\% with a bias of $72$ V for the optical power of $138$ nW. It is important to point out again that EQE exceeding $100$\% indicates that the number of generated photo-carriers is more than the total number of incident photons on the device, which is strong evidence of carrier \textit{avalanche multiplication} taking place in the high electric-field region in the Si APD. It should be noted that the GeSn-on-Si APD has only $25$ nm of active material thickness that is responsible for the absorption of the $1.3\,\mu$m laser. This is likely the reason behind the consistently global EQE reduction as the incident optical power increases, and it strongly indicates a sign of saturation in the total current that is dominated by photo-generated carriers, as shown in Fig. S1(b).

\subsection {\bf S2. The influence of varying device diameter and contacts annealing on the breakdown characteristics of the Si and GeSn-on-Si APD devices.}

\medskip

\begin{figure*}
    \centering
    \includegraphics[scale=0.62]{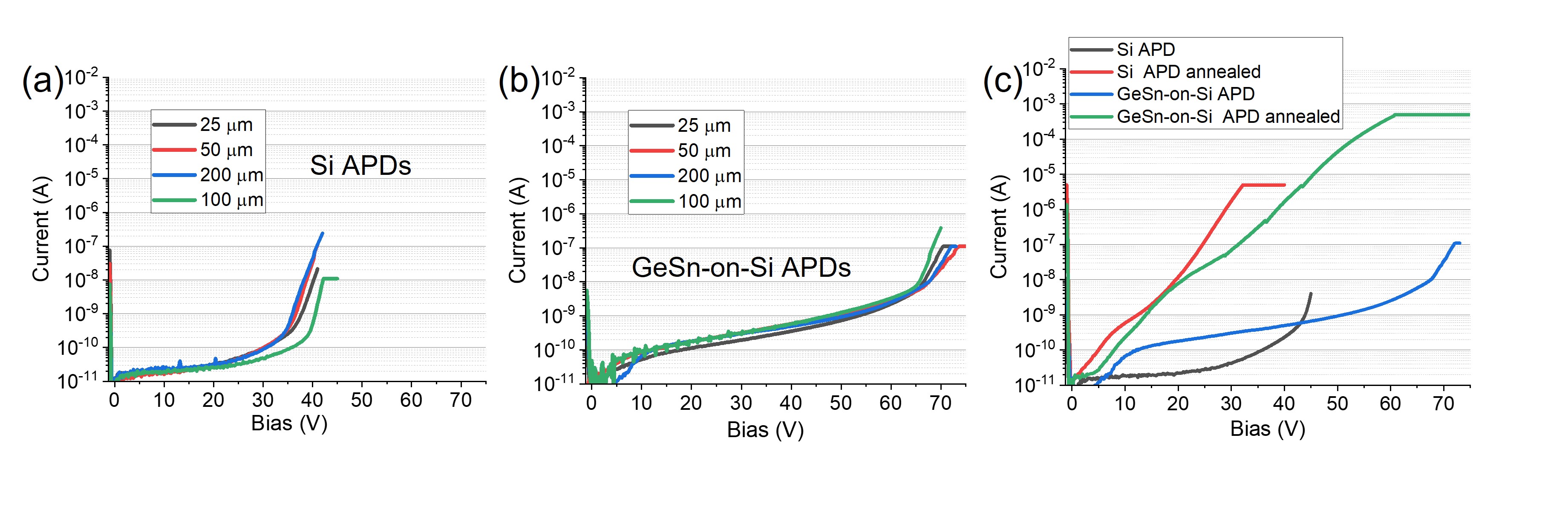}
    \caption{{\bf LT I-V curves of GeSn-on-Si APDs in dark illumination condition and additional contacts annealing.} (a) I-V curves for the Si APD device measured at $78$ K in dark for various device diameters. (b) same as (a) but for the GeSn-on-Si devices.(c) Additional contacts annealing was made for the Si and GeSn-on-Si APDs at $385 \,^{\circ}\text{C}$ for $15$ seconds.}
\end{figure*}

\medskip

Despite the fact that the breakdown properties are more relevant to characterization of a single-photon avalanche detector (SPAD), we have studied the breakdown of the device herein for the sake of completeness and to find out if these APDs can operate as SPADs.
It is important to mention that the breakdown voltages of these GeSn-on-Si APDs were found to be independent of the device diameter, as well, which indicates that they are not influenced by the device geometry but by the high electric field induced at the multiplication region inside the Si substrate. Since the p-n region inside the Si substrate had the same cross-sectional width, doping level, and annealing conditions, the highest electric field formed was independent of device diameter geometry and device breakdown as well, as shown in Figs. S2(a) and S2(b). 

In addition to the device diameter comparison and its influence on the breakdown characteristics, a contacts annealing was attempted to hopefully reduce contact resistance and reduce the on/off response delay of the device. The Si and GeSn-on-Si devices were annealed using rapid thermal annealing (RTA) at a temperature of $385 \,^{\circ}\text{C}$ for $15$ seconds, and the IV curves were remeasured and compared with those before annealing, as shown in Fig. S2(c). In general, the annealing step has caused significant leakage in both devices, which pushed the devices to the current compliance limit before showing a steep breakdown characteristic, unlike the case before annealing.

\subsection {\bf S3. Assessing the feasibility of measuring single photons using GeSn-on-Si APDs.}

\medskip

\begin{figure*}
    \centering
    \includegraphics[scale=0.75]{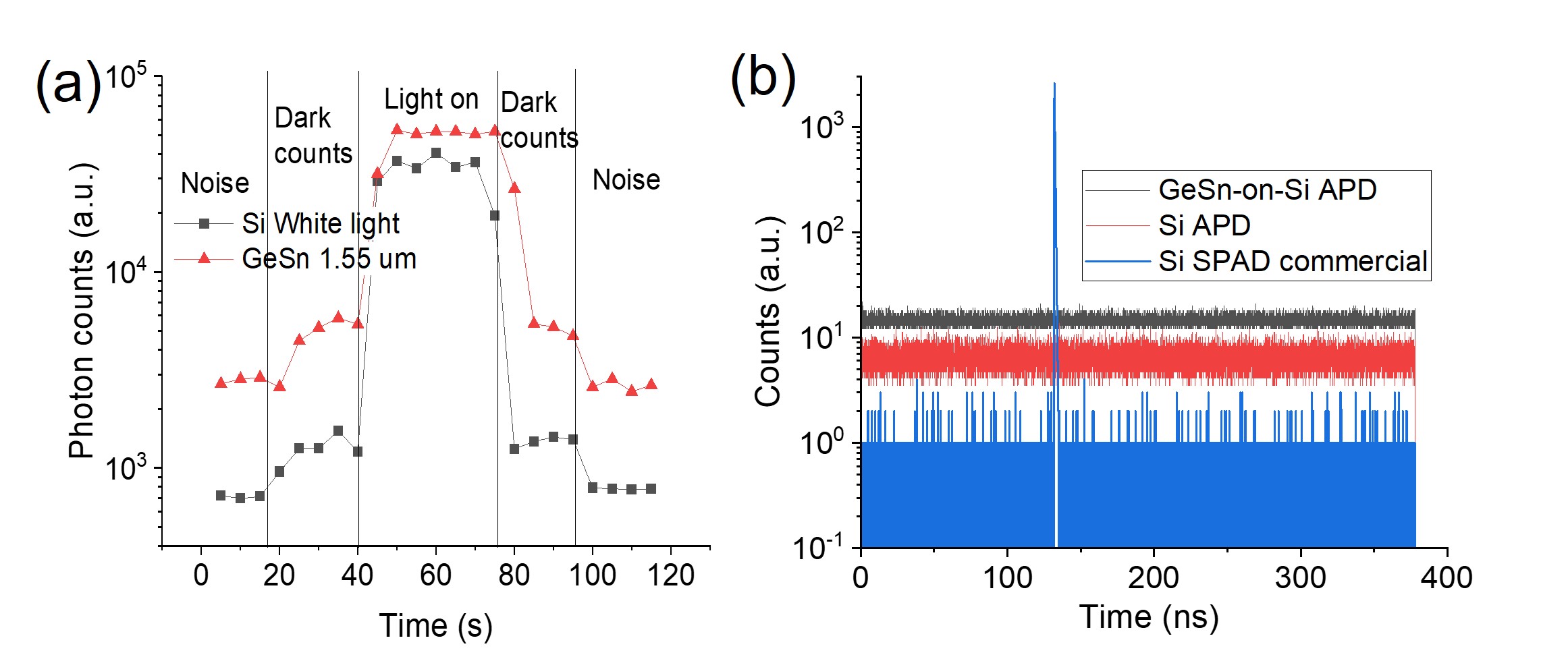}
    \caption{{\bf LT measurements of single photon counting using GeSn-on-Si APDs under $1.55 \,\mu$m illumination at the breakdown voltage.} (a) Free running single photon measurements, and it shows single photon counts form the fabricated Si and GeSn-on-Si devices. (b) Gated single photon signal measurements for commercial Si SPAD compared to that of the Si and GeSn-on-Si devices at 2.5 MHz frequency. No gated single photon signal was detected by the fabricated devices. }
\end{figure*}

\medskip

Since these APDs devices maintained a relatively low dark current until breakdown, especially for the Si APDs, it was imperative to investigate the single-photon operation-mode of these Si and GeSn-on-Si devices. The setup was configured as a gated single-photon measurement, where the SPAD device was biased (gated) above breakdown for a short period of time ($\sim 150$ ns) and then returned to below breakdown bias with a gate pulse frequency of 2.5 MHz. This gated measurement configuration helps to avoid overwhelming the SPAD with dark counts, since the SPAD is off most of the time, which is useful for the GeSn-on-Si device in comparison to the Si one. \cite{vines2019high} A picosecond pulse laser was synchronized using a master-clock to the same gate frequency to repeatedly incident light when the SPAD gate pulse is high. Two picosecond lasers with wavelengths $908$ nm and $1.3 \,\mu$m were used in these measurements of the Si and GeSn-on-Si devices, respectively. The device under test was placed in the cryoprobe station at $78$ K and the laser light was incident using an optical fiber embedded inside the cryoprobe station. A bias tee was used to simultaneously provide the DC bias and gate pulse to the measured device. Additionally, a PicoQuant time-correlated single-photon counter was used to count and record avalanche breakdown events using an integrated current comparator and an accurate time stamp recorder. 

In Fig. S3(a), the device counts are plotted as a function of time when white light is incident on the Si device and when a continuous wave $1.55 \,\mu$m laser is incident on the GeSn-on-Si device. The noise mentioned in this figure is the random counts coming for the electronics while the device under test is not biased, the dark counts are the counts that the TCSPC measures when the device is biased close to the breakdown voltage, and the light-on counts are the counts that the TCSPC measures when the light is on and the device is under bias. It is clear from Fig. S3(a) that both Si and GeSn-on-Si devices indicate single-photon counts measured by the TCSPC. However, upon utilizing a synchronized picosecond laser, no signals were observed in the TCSPC histogram of the Si and GeSn-on-Si devices, unlike the case of a commercial Si SPAD measured using the same setup, as shown in Fig. S3(b). In Fig. S3(b), the histograms of the measured device counts are plotted as a function of the time interval period which is the inverse of the gate pulse frequency. This is most likely because of the reduced bandwidth and high latency of the fabricated Si and GeSn-on-Si devices, owing to the high contacts resistance and device capacitance that act as a low pass frequency filter and prevent the devices from synchronizing with the incident photon events at specific periodic time stamp.

\bibliography{SImain}